\documentclass[prd,aps,twocolumn,amsmath,amssymb,nofootinbib,preprintnumbers]
{revtex4}

\usepackage{graphicx}
\usepackage{dcolumn}
\usepackage{bm}
\usepackage{amsmath}
\usepackage{amsthm}

\usepackage{amsfonts}
\usepackage{euscript,bbm}
\usepackage{ifthen}
\usepackage{psfrag}
\usepackage{slashed}

\newtheorem{cond}{Condition}[section]
\newtheorem{nogo}{No-Go}[section]

\def\ls{\mathrel{\lower4pt\vbox{\lineskip=0pt\baselineskip=0pt
           \hbox{$<$}\hbox{$\sim$}}}}
\def\gs{\mathrel{\lower4pt\vbox{\lineskip=0pt\baselineskip=0pt
           \hbox{$>$}\hbox{$\sim$}}}}
\def\drawbox#1#2{\hrule height#2pt

\hbox{\vrule width#2pt height#1pt \kern#1pt
              \vrule width#2pt}
              \hrule height#2pt}

\def\Asym#1#2{\vcenter{\vbox{\drawbox{#1}{#2}
              \kern-#2pt       
              \drawbox{#1}{#2}}}}

\newcommand{\be}{\begin{equation}}
\newcommand{\ee}{\end{equation}}
\newcommand{\bea}{\begin{eqnarray}}
\newcommand{\eea}{\end{eqnarray}}

\newcommand{\ibar}{\overline{i}}
\newcommand{\jbar}{\overline{j}}
\newcommand{\kbar}{\overline{k}}
\newcommand{\lbar}{\overline{l}}

\newcommand{\nbar}{\overline{n}}
\newcommand{\Phibar}{\overline{\Phi}}
\newcommand{\Psibar}{\overline{\Psi}}

\newcommand{\Qbar}{\overline{Q}}
\newcommand{\p}{\partial}

\begin{document}

%
\title{Holomorphic Bisectional Curvatures, Supersymmetry Breaking, and Affleck-Dine Baryogenesis}

\author{Bhaskar Dutta$^{1}$}
\author{Kuver Sinha$^{1}$}

\affiliation{$^{1}$~Department of Physics, Texas A\&M University, College Station, TX 77843-4242, USA \\
}

\begin{abstract}




Working in $D=4, N=1$ supergravity, we utilize relations between holomorphic sectional and bisectional curvatures of Kahler manifolds to constrain Affleck-Dine baryogenesis. We show the following No-Go result: Affleck-Dine baryogenesis cannot be performed if the holomorphic sectional curvature at the origin is isotropic in tangent space; as a special case, this rules out spaces of constant holomorphic sectional curvature (defined in the above sense) and in particular maximally symmetric coset spaces. We also investigate scenarios where inflationary supersymmetry breaking is identified with the supersymmetry breaking responsible for mass splitting in the visible sector, using conditions of sequestering to constrain manifolds where inflation can be performed.

\end{abstract}
MIFPA-12-18
\maketitle


\section{Introduction}

Supersymmetry, if it is a symmetry of nature, had important consequences for the physics of the very early universe. This can be understood at several different levels. One of the most fundamental connections between supersymmetry and early universe cosmology lies in the fact that the finite energy density during inflation \cite{Guth:1980zm} breaks supersymmetry. This breaking of supersymmetry induces soft terms along flat directions in the scalar potential, the existence of which constitutes another central feature of supersymmetric (as opposed to non-supersymmetric) theories. 


The first important effect of supersymmetry in this context is the production of condensates of non-relativistic particles. The analysis starts with the observation that the inflation-induced soft terms (usually called the Hubble-induced terms) are of order the Hubble scale $H$, and in the early universe, during inflation, this is typically taken to be
\be
H \, \gg \, m_{3/2} \,\,.
\ee
In the above, $m_{3/2}$ gives the scale of hidden sector supersymmetry breaking that gives the mass splitting of particles and superpartners in our universe. The fields are thus critically (and not over) damped and follow the instantaneous minimum of the potential. If initial conditions are set up such that the instantaneous minimum is displaced from the true minimum (as happens when the Hubble-induced and hidden sector-induced soft masses have opposite sign), then the field begins to oscillate when $H$ becomes $\mathcal{O}(m_{3/2})$. These oscillations correspond to a condensate of non-relativistic particles, and can have dramatic consequences on the history of the universe. 

Thus, the magnitude and sign of inflation-induced soft masses clarify initial conditions for paradigms such as Affleck-Dine baryogenesis \cite{Dine:1995uk}, in addition to providing a clearer picture of the cosmological moduli problem.

The second important effect of supersymmetry may simply be to set the scale of early universe physics. This was observed by Kallosh and Linde \cite{Kallosh:2004yh}, by studying the effect of inflation on the stabilized potential of moduli in $D=4, N=1$ effective supergravity. Finite vacuum energy $V_0$ in some sector (the energy during inflation provides the most compelling example) induces a potential of the form $\sim V_0/\sigma^n$ along a modulus $\sigma$, from Weyl rescaling. This induced potential may destabilize the moduli; the importance of this effect clearly depends on the height of the barrier protecting the moduli. In the most widely studied example of KKLT \cite{Kachru:2003aw}, the barrier is of $\mathcal{O}(m_{3/2})$, thus providing a critical value of the vacuum energy during inflation $H \, \sim \, m_{3/2}$. This suggests that it may be interesting to study inflation and supersymmetry breaking as related paradigms - for example, one can ask: what kinds of effects do slow-roll inflation leave on soft masses? Or, going the other way, how can known properties of soft masses influence inflation? For example, one way to address the flavor problem is sequestering, which places constraints on the allowed underlying Kahler geometry; could such constraints back-react on inflation?

Clearly, the setting for understanding the effects of supersymmetry in the early universe is $D=4, N=1$ supergravity. In the case of the production of condensates, this is necessitated by the importance of Planck scale operators that give rise to important couplings between the supersymmetry breaking field and the flat direction. In the second case mentioned above, the potential for string moduli are usually studied in the effective $D=4, N=1$ supergravity Lagrangian, that comes from a suitable string compactification.

Given the above, can one make general statements about when Affleck-Dine baryogenesis would be possible, or, in the other case, where properties of soft masses may be in conflict with inflation? This depends on how general one wants to be - so for example, making statements about the superpotential $W$ may be more model-dependent but robust, given that the superpotential is protected by holomorphy and symmetries. On the other hand, making statements about the Kahler potential could be powerful enough to rule out classes of manifolds where these phenomena can take place, based on geometry alone.

In this paper, we will follow the route of looking at these questions purely from the point of view of geometry. We will ask: can one make broad statements about the underlying Kahler geometry of $D=4, N=1$ effective supergravity, in relation to Affleck-Dine baryogenesis and inflation-induced soft masses? 

This will necessitate using several necessary conditions for the various paradigms mentioned above in terms of quantities that are fundamental to the underlying Kahler geometry, without relying on information about the superpotential. This involves two aspects: necessary geometric conditions for inflation to happen, and necessary conditions for induced scalar masses to satisfy a requisite feature (say a condition to obtain a tachyonic mass). 

While such conditions could be given at various levels of sophistication, we will choose the simplest conditions available. On the inflation side, the relevant condition will involve the holomorphic sectional curvature of the plane (in tangent space) spanned by the field breaking supersymmetry (say $(\Psi, \Psibar)$). These conditions have been worked out in great detail by \cite{Covi:2008cn}. We note that these conditions apply for $F-$term inflationary models; we will be restricting ourselves to this class for the purposes of this paper, reserving the case of $D-$term inflation for future work.

For the scalar masses, the relevant conditions are best expressed in terms of the holomorphic bisectional curvature \cite{GoldbergKob} of the $(\Psi, \Psibar)$ plane and the plane spanned by the scalar field $(Q, \Qbar)$ one is interested in studying. The use of the holomorphic bisectional curvature is already implicit in \cite{Kaplunovsky:1993rd}, and has been recently stressed in \cite{Marsh:2011ud}, \cite{Dutta:2010sg}.

 A summary of these results is provided in Table \ref{HoloBiholoresults}. As we go along the text, we will provide details for each entry.

Secondly, having expressed our conditions in terms of these quantities, we will stress that one can go on to constrain scenarios where inflation-induced soft masses play a role by exploiting relations between holomorphic sectional and bisectional curvatures on classes of manifolds. Again, how general these constraints are would depend on how precisely one wants to pinpoint a manifold. 

For example, we will prove the following No-Go Theorem: \textit{Affleck-Dine baryogenesis cannot be performed if the holomorphic sectional curvature at the origin (or whatever vev fields are expanded about) is constant for all choices of planes in the tangent space. In particular, this rules out Kahler manifolds with constant holomorphic sectional curvatures (in the above sense), and as a special case, maximally symmetric coset spaces}.

While we are agnostic about the UV completion of our effective supergravity setting, we will note that this rules out, as special cases, the simplest coset spaces available from type IIB compactifications, a fact already noted in \cite{Marsh:2011ud}, \cite{Dutta:2010sg}.

Likewise, we will continue our investigation into scenarios where one wants to constrain inflationary physics from flavor physics, using geometric conditions from sequestering. We will see that the simplest sequestering geometries, which have constant holomorphic sectional curvature, also cannot accomodate inflation. We will briefly look at more complicated sequestering geometries, which satisfy the necessary condition for inflation.

\begin{table}[!htp] 
\begin{center}
\begin{tabular}{c c c c } \hline \hline
\\
Phenomenon          & \,\,\, $\mathbb{H}[\Psi]$ \,\,\,  & \,\,\,$\mathbb{H}[\Psi,Q]$ \,\,\,  & \,\,\,Comment\,\,\,  \\ \\ \hline \hline

\\
Inflation           &   $> \, -\frac{2}{3}$   & -                     &  Necessary \\ \\
A-D Baryogenesis    &  -                    &  $< \, -1$             &  Necessary \\ \\
Sequestering        &  -                    & $= \, - \frac{1}{3}$    &           Sufficient \\

\\ \hline \hline
\end{tabular}
\end{center}
\caption{Conditions on the holomorphic sectional and bisectional curvatures of the Kahler manifold in $D=4,N=1$ supergravity for different phenomena. $\mathbb{H}[\Psi]$ is the holomorphic sectional curvature of the plane spanned by the field breaking supersymmetry $(\Psi, \Psibar)$. $\mathbb{H}[\Psi, Q]$ is the holomorphic bisectional curvature of the $(\Psi, \Psibar)$ plane and the $(Q, \Qbar)$ plane, where $Q$ denotes a supersymmetric flat direction in the case of Afflck-Dine baryogenesis, and a generic matter field in the case of sequestering. We follow the conventions of \cite{GoldbergKob}. Details in the text.}
\label{HoloBiholoresults}
\end{table}

The plan of the paper is as follows. In Section \ref{secbisect}, we will define holomorphic sectional and bisectional curvatures in the context of $D=4, N=1$ supergravity and state certain relations between them that will be of use to us. In Section \ref{inflationsection}, we show a condition on holomorphic sectional curvatures that is necessary for inflation. In Section \ref{bisectsoftmasses}, we write expressions for the soft masses in effective supergravity in terms of holomorphic bisectional curvatures, in a form that is most useful for us. In Section \ref{ADbarsection}, we consider Affleck-Dine baryogenesis, showing a necessary condition on holomorphic bisectional curvatures and a No-Go result. In Section \ref{infsusybreak}, we consider the scenario when the supersymmetry breaking during inflation is identified as the sector responsible for mass splitting in the visible sector, studying sequestering in this context.






\section{Holomorphic Sectional and Bisectional Curvatures} \label{secbisect}

In this Section, we write expressions for the holomorphic sectional and bisectional curvatures in terms of the fields appearing in the $D=4, N=1$ supergravity potential. Then, we mention relations between them that we will find useful in the rest of the paper. For more careful definitions of all quantities and the proofs of the relations, we refer to the Appendix.

We consider chiral multiplets $\Phi_I \equiv (\Phi_i,\Phi_{\ibar})$ (henceforth, we will use the same notation for a superfield and its scalar component). The potential for the scalar components of the $\Phi_I$ depends on the Kahler potential $K$ and the superpotential $W$, and their derivatives with respect to $\Phi_i$ and $\overline{\Phi}_{\overline{j}}$.

We will not consider any particular UV completion; rather, we will view our setting as an effective theory, with a superpotential and Kahler potential determined by physics at a higher scale. Among the set of fields $\{\Phi_I \}$, there is a set of SUSY breaking fields $\{ \Psi_I \} \subset \{\Phi_I \}$ that acquire non-zero $F-$terms thus breaking supersymmetry during inflation or afterwards. There are also other fields $\{Q_I\} \subset \{\Phi_I\}$, which constitute chiral fields that may be flat directions in the visible sector, or generic fields, depending on the context.

$K$ defines the Kahler potential of a Kahler manifold spanned by the $(\Phi_i,\Phibar_{\ibar})$. The potential for the scalar fields $\Phi_I$ depends on the Kahler geometry as well as the superpotential $W$, and is given by
\be
V \, = \,  F_iF_{\jbar}g^{i\jbar} - 3m^2_{3/2} \,\,,
\ee
where 
\bea
F_i  &=&  e^K D_i W = e^K(\p_{i}W + W\p_{i}K)\nonumber \\
m_{3/2}^2 &=& e^K |W|^2 \,\,.  
\eea

Since the $\Psi^I = (\Psi^i, \Psi^{\overline{i}})$ define the SUSY breaking direction in field space, one has
\be
F^{\Psi_I} \, \neq \, 0 \,\,\,.
\ee

The metric, connection, and curvature tensor of the Kahler manifold are given by
\bea
g_{i\jbar} &=& \partial_{i}\partial_{\jbar} K \nonumber \\
\Gamma_{ij}^k &=& g^{\lbar k}\partial{i}g_{j\lbar} \nonumber \\
R_{\kbar l \ibar j} &=& \p_{\ibar} \p_{j} \p_{l} \p_{\kbar} K \, - \, g^{\nbar m}(\p_{\ibar} \p_{\kbar} \p_{m} K)(\p_{j} \p_{l} \p_{\nbar} K) \,\,\,.
\eea

The holomorphic sectional curvature of a plane $(\Psi, \Psibar)$ (defined in the tangent space at a given point in field space) is defined in components by
\be \label{sectcurvcplx}
\mathbb{H}[\Psi] \, = \, - \frac{R_{\Psi \Psibar \Psi \Psibar}}{g_{\Psi \Psibar} g_{\Psi \Psibar}} \,\,.
\ee
We will be following the conventions of \cite{GoldbergKob} throughout, note in particular the negative sign in the above equation. 

We will also be interested in the holomorphic bisectional curvature of a plane $(\Psi, \Psibar)$ and a plane $(Q, \Qbar)$. We will assume that the fields $Q$ have been rescaled such that $g_{Q \Qbar} = 1$, as is usually done for the matter fields, and that the planes defined by $\Psi$ and $Q$ are orthogonal. Thus, we will assume $g_{Q,\Psibar} = 0$, which is usually the case when one studies the potential near the origin of $Q$. One then has
\be
\mathbb{H}[\Psi, Q] \, = \, - \frac{R_{\Psi \Psibar Q \Qbar}}{g_{\Psi \Psibar}}\,\,.
\ee
Although we write the metric $g_{\Psi \Psibar}$ in the above definitions, we will usually normalize the field direction.

 

\subsection{Relations between Holomorphic Sectional and Bisectional Curvatures}

We will be particularly interested in relations between holomorphic sectional and bisectional curvatures. The reason should be clear from Table I: the necessary condition for inflation that we will be showing in the next Section is a condition on the holomorphic sectional curvature $\mathbb{H}[\Psi]$, while the conditions on the inflation-induced masses, which we will be showing in Section \ref{bisectsoftmasses}, are conditions on the holomorphic bisectional curvature $\mathbb{H}[\Psi, Q]$. 

The relation we will use is the following. At a given point $x$ in the manifold, for orthonormal directions $Q$ and $\Psi$, the holomorphic bisectional curvature $\mathbb{H}[Q,\Psi]$ between the planes $(Q,\Qbar)$ and $(\Psi,\Psibar)$  may be written as a linear combination of certain holomorphic sectional curvatures:
\be \label{associatedrelation}
\mathbb{H}[\Psi, Q] \, = \, \frac{1}{4}\{\sum_{a=1}^{4}\mathbb{H}[\lambda_a] -\mathbb{H}[\Psi] - \mathbb{H}[Q]\}\,\,,
\ee
where the $\lambda_a$ denote certain holomorphic and anti-holomorphic sections associated with the section spanned by the pair $(Q, \Psi)$.

In order to demonstrate the utility of this relation, we will focus, in this paper, on the special case where the holomorphic sectional curvatures are simply constant for all choices of planes in tangent space at $x$. 
\be
R_{j\jbar j \jbar} = {\rm constant} \,\,\, \forall \,\,\, [{\rm span}(\partial_j,\partial_{\jbar}) \in T_{x}(\mathcal{M})] \,\,.
\ee
In that case, one obtains
\bea \label{constrelation}
&&\mathbb{H}[\Psi] = {\rm const. \,\, } (c) \nonumber \\
&\Rightarrow & \,\, \frac{|c|}{2} \, \leq \, |\mathbb{H}[\Psi, Q]| \, \leq \, |c| \,\,.
\eea
For orthonormal planes, the lower bound is exactly satisfied, as follows from Eq.~\ref{associatedrelation}. The general inequality will be shown in the Appendix.

If the isotropy of the holomorphic sectional curvature in tangent space holds for all $x$ belonging to the Kahler manifold, we say that the manifold has constant holomorphic sectional curvature. We note that this is a statement about special components of the curvature tensor; namely, a manifold has constant holomorphic sectional curvature when
\be
R_{j\jbar j \jbar} = {\rm constant} \,\,\, \forall \,\,\, [x \in \mathcal{M}, {\rm span}(j\jbar) \in T_{x}(\mathcal{M})] \,\,.
\ee

Clearly, for manifolds of constant holomorphic sectional curvature $c$, the holomorphic bisectional curvature is bounded between $c/2$ and $c$. This fact can be proven without recourse to  Eq.~\ref{associatedrelation}, and we will do so in the Appendix. As a special case, one can moreover consider the case when the full curvature tensor is also covariantly constant, apart from being isotropic in tangent space at all points. For example one has the maximally symmetric coset spaces, for which the above bound holds, as can be checked explicitly starting from the Kahler metric of such spaces.

In the next Section, we will derive the necessary condition for inflation that we will be interested in. We will then have two applications in mind. First, we will consider the case of Affleck-Dine baryogenesis, where there are certain requirements on the vacuum-energy induced soft mass along a baryon number carrying supersymmetric flat direction. Next, we will consider the case where the supersymmetry breaking during inflation is taken as the breaking that is responsible for mass splitting in the visible sector.


\section{Necessary Conditions for Inflation} \label{inflationsection}

The vacuum energy during inflation breaks supersymmetry; we assume that the field $\Psi$ is responsible for the supersymmetry breaking. In this section, following \cite{Covi:2008cn}, we prove a necessary condition on the holomorphic sectional curvature of the $(\Psi, \Psibar)$ plane for inflation to take place. This proves the first entry in Table I. 

Inflation requires that certain slow-roll parameters which depend on the derivatives of the scalar potential be small. In a general manifold, the slow-roll parameters are given by \cite{Burgess:2004kv}
\bea
\epsilon \, &= \, \frac{\nabla^i V \nabla_i V}{V^2} \nonumber \\
\eta \, &= \, {\rm min \,\,\, eigenvalue} \, \{ N \} \,\,,
\eea
where 
\be \label{Hessian}
N \, = \, \frac{1}{V} \left( \begin{array}{cc}
\nabla^i \nabla_j V  &  \nabla^i \nabla_{\overline{j}} V  \\
\nabla^{\overline{i}}\nabla_j V & \nabla^{\overline{i}} \nabla_{\overline{j}}V   
\end{array} \right) \,\,.
\ee
In the above, we have used the covariant derivative on the Kahler manifold $\mathcal{M}$
\be
\nabla_{i}f^k \equiv \p_{i}f^k + \Gamma_{ij}^k f^j  
\ee
for any vector $f^k$ on $\mathcal{M}$. In the above, $I = (i, \overline{i})$ and $J = (j, \overline{j})$ and $\nabla_i$ is a covariant derivative with respect to the metric $g_{i \overline{j}}$. 

These expressions for the slow-roll parameters reduce to the usual ones in the Gaussian normal frame. 

Since $\eta$ is defined as the minimum eigenvalue of the matrix $N$, it always satisfies a bound. For any given unit vector $u^I = (u^i,u^{\ibar})$ one has
\be
\eta \, \leq \, u_I N^I_J u^J \,\,.
\ee

The goal is to extract from here a necessary condition that is dependent on the choice of the Kahler geometry of $\mathcal{M}$, but independent of specific choices of the superpotential. The correct choice of $u^I$ turns out to be
\be
u^I = (F^{\Psi}, F^{\Psibar})/(\sqrt{2}|F|)\,\,,
\ee
that is, the normalized SUSY breaking direction.

Evaluating the relevant covariant derivatives, the bound turns out to be
\be
\eta \, \leq \, \eta_{\rm max} \, \equiv \,  \frac{2}{3\gamma} + \frac{1+\gamma}{\gamma}\mathbb{H}[\Psi] + \mathcal{O}(\sqrt{\epsilon}) \,\,,
\ee
where
\be
\gamma \, = \, \frac{1}{3}\frac{V}{m^2_{3/2}} \, \sim \, \frac{H^2}{m^2_{3/2}} \,\,.
\ee

Here, $\mathbb{H}[\Psi]$ is the holomorphic sectional curvature along the SUSY breaking plane defined by $(\Psi, \overline{\Psi})$. It has been defined in terms of the curvature tensor on the Kahler manifold in Eq. \ref{sectcurvcplx}.

We drop all terms involving $\epsilon$, since $\sqrt{\epsilon} < \mathcal{O}(10^{-3})$. Now, the spectral index is given by
\be
n_s \, = \, 1+2\eta \,\, \Rightarrow \,\, \eta_{\rm observed} \sim -0.01\,\,\,\,.
\ee
Therefore, $\eta_{\rm max} \geq -0.01$.

This yields the following necessary bound on the holomorphic sectional curvature along the SUSY breaking direction
\be \label{inflationfullcond}
\mathbb{H}[\Psi] \, > \, - \frac{2}{3}\frac{1}{1+\gamma} \,\,.
\ee

While this bound depends on the ratio of the inflationary scale and the value of $m_{3/2}$, there is a hard bound  \cite{Covi:2008cn}

%
%
%

\begin{cond} \label{conditioninflation}
Inflation is only possible on manifolds $\mathcal{M}$ where the holomorphic sectional curvature along the SUSY breaking direction satisfies 
\be
\mathbb{H}[\Psi] \, > \, - \frac{2}{3} \,\,.
\ee
This is a necessary (but not sufficient) condition.
\end{cond}

Clearly, the condition is not sufficient - obtaining the correct slow-roll parameters required for inflation needs other non-trivial conditions, including a full understanding of the superpotential, and there is no guarantee that inflation is even possible on a given manifold. For example, flat manifolds with canonical Kahler potential, which form the setting of many studies of inflation, trivially satisfy the above bound, but nonetheless setting up inflation requires non-trivial fine-tuning.

\section{Holomorphic Bisectional Curvatures and Soft Masses} \label{bisectsoftmasses}

In this section, we recall the expression for the soft masses for chiral superfields in supergravity, in particular writing them in terms of holomorphic bisectional curvatures. These are general expressions with no assumption about the supersymmetry breaking sector. For us, this means that the supersymmetry breaking considered in this section does not necessarily have anything to do with inflation, although we will continue to use the same notation as before.

Masses are most easily obtained from the Hessian matrix in Eq.~(\ref{Hessian}). In particular, one obtains
\bea \label{massnabla}
&& \nabla_i \nabla_{\jbar} V = (m_{3/2}^2 + V_0)g_{i\jbar} - R_{i\jbar k \lbar}F^{k}F^{\lbar} \nonumber \\
&+& ({\rm terms \,\, proportional\,\, to\,\,} F_i, F_{\jbar}, \nabla_i F_k, \nabla_{\jbar}F_{\lbar})
\eea
We will consider masses for a chiral superfield $(i,\ibar) \equiv (Q,\overline{Q})$, which does not participate in supersymmetry breaking. Supersymmetry breaking is dominated by a field $(\Psi,\Psibar)$. The terms in the second line all vanish (covariant derivatives like $\nabla_{Q}F_{\Psi}$ vanish by extremising $V$ with respect to $Q$). We normalise the fields $Q$ such that $g_{Q \Qbar} = 1$. Moreover, we use the fact that
\be \label{V_0forPsi}
F^{\Psi}F^{\overline{\Psi}}g_{\Psi \overline{\Psi}} \, = \, V_0 + 3 m_{3/2}^2 \,\,.
\ee
%

%
%

Using Eq.~(\ref{massnabla}) and Eq.~(\ref{V_0forPsi}), we therefore have
\be \label{formulaformasses}
m_{Q \Qbar}^2 \,\, = \,\, V_0(1 + \mathbb{H}[Q,\Psi]) \, + \, 3m_{3/2}^2(\frac{1}{3} + \mathbb{H}[Q,\Psi])\,\,. 
\ee

Eq.~(\ref{formulaformasses}) represents the most transparent expression of soft masses in terms of the holomorphic bisectional curvature. 

At this point, we can distinguish between two scenarios. 

First: during a period of vacuum energy domination such as inflation, supersymmetry breaking is dominated by a field $\Psi_1$, which continues to be the dominant source of supersymmetry breaking responsible for mass splittings in the visible sector, in the final vacuum of the theory with negligible vacuum energy. We denote the soft mass of $Q$ induced by supersymmetry breaking during inflation as $m_{\rm soft,inf}$, and the soft mass induced by supersymmetry breaking in the final vacuum by $m_{\rm soft,final}$.
%
%
\bea
&&\;\;\;\;\,\,\,\,\;\;\;{\rm During \,\, inflation:} \nonumber \\
m_{\rm soft,inf}^2  &=&  V_0(1 + \mathbb{H}[Q,\Psi_1])  +  3m_{3/2}^2(\frac{1}{3} + \mathbb{H}[Q,\Psi_1]) \nonumber 
\eea
\bea \label{firstscenario}
&&{\rm Final \,\, vacuum:} \nonumber \\
m_{\rm soft,final}^2  &=& 3m_{3/2}^2(\frac{1}{3} + \mathbb{H}[Q,\Psi_1])
\eea

Second: the field $\Psi_1$ dominates the supersymmetry breaking during inflation, but contributes negligibly to supersymmetry breaking in the final vacuum where the vacuum energy is negligible. Another field $\Psi_2$ dominates supersymmetry breaking in the final vacuum. It is typical to assume in this case that the scale of inflation is much larger than $m_{3/2}$.
\bea \label{massestwofields}
&&\;\;{\rm During \,\, inflation:} \nonumber \\
&& m_{\rm soft,inf}^2  =  V_0(1 + \mathbb{H}[Q,\Psi_1]) \nonumber 
\eea
\bea
&&{\rm Final \,\, vacuum:} \nonumber \\
m_{\rm soft,final}^2  &=& 3m_{3/2}^2(\frac{1}{3} + \mathbb{H}[Q,\Psi_2])
\eea

We distinguish between these two scenarios because of the applications we have in mind. In the next Section, we discuss Affleck-Dine baryogenesis, in which we will assume that the second scenario is operational. Subsequently, we will study the former scenario, which assumes a matching of inflationary and supersymmetry breaking scales.


\section{Affleck-Dine Baryogenesis} \label{ADbarsection}
 
Affleck-Dine baryogenesis \cite{Dine:1995uk} relies on the vacuum energy during an inflationary era to produce coherent oscillations along a supersymmetric flat direction. The interaction between the inflationary sector and the flat direction occurs through Planck-suppressed operators.

The normalized scalar component of a composite gauge invariant product of chiral superfields with non-zero net baryon or lepton number constitutes a suitable flat direction. For example, for $H_uL$, the flat direction $Q$ is given by
\be
H_u \, = \, \frac{1}{\sqrt{2}}\left(
\begin{matrix}
0 \\
Q \\
\end{matrix}
\right )\, ,
\,\,\,\,
L \, = \, \frac{1}{\sqrt{2}}\left(
\begin{matrix}
Q \\
0 \\
\end{matrix}
\right )
\ee
If the flat direction $Q$ is initially displaced from its true minimum at the origin during inflation, it subsequently oscillates when $V_0$ becomes smaller than the effective mass which is $\sim m_{3/2}$. The energy of the oscillations corresponds to a condensate of non-relativistic particles. A net baryon asymmetry may be produced during oscillation depending on the magnitude of baryon number-violating terms in $V(Q)$.

Schematically, one has the following. Supposing that flat directions are lifted by non-renormalizable terms in the superpotential
\be
W \, = \, \frac{\lambda}{nM_P^{n-3}}\phi ^n \,\, ,
\ee
the potential along $Q$, taking into account supersymmetry breaking terms due to the finite energy during inflation, is written as
\bea
V(Q) &=& (m_{\rm soft,inf}^2 + m_{\rm soft,final}^2)|Q|^2 + \left( \frac{(A+a_{\rm inf})\lambda Q ^n}{n M_P^{n-3}} + {\rm h.c.} \right) \nonumber \\
&+& |\lambda|^2 \frac{|Q | ^{2n-2}}{M_P^{2n-6}} \,\, .
\eea
Here, $m_{\rm soft,inf}^2$ and $a_{\rm inf}$ denote soft parameters induced by supersymmetry breaking during inflation, while $m_{\rm soft,final}$ and $A$ arise from supersymmetry breaking sector at the end of inflation, in the final vacuum of the theory. 

Clearly, if $m_{\rm soft,inf}$ is tachyonic, the field $Q$ acquires a non-zero vacuum expectation value during inflation. It thereafter remains critically damped, and tracks an instantaneous minimum as long as $m_{\rm soft,inf}^2 \gg  m_{\rm soft,final}^2$, which is roughly equivalent to $H \gg m_{3/2}$, from Eq.~(\ref{massestwofields}). The minimum lies at
\be
|Q | \sim \left(\frac{\sqrt{-c}H M_P^{n-3}}{(n-1)\lambda}\right)^{\frac{1}{n-2}}
\ee
and the field tracks this minimum until $H \sim m_{3/2}$. The inclusion of the Hubble-induced A-term $a_HH$ results in $n$ discrete vacua in the phase of $Q$, and the field settles into one of them. When $H \sim m_{3/2}$ the field begins to oscillate around the new minimum $Q = 0$; thereafter the soft $A$-term becomes important and the field obtains a motion in the angular direction to settle into a new phase. The baryon number violation thus becomes maximal during this time and imparts asymmetry to the condensate. The final baryon to entropy ratio depends on the resulting baryon number per condensate particle, the total energy density in the condensate, and the inflaton reheat temperature.

We will take the mass relation Eq.~(\ref{massestwofields}) as a starting point, thereby assuming that the inflationary supersymmetry breaking at high scale is due to $\Psi_1$, and the final supersymmetry breaking is due to $\Psi_2$.

%
%





From 
\be
m_{\rm soft,inf}^2 \,\, = \,\, V_0(1 + \mathbb{H}[Q,\Psi]) \,\, < \,\, 0 \,\,,
\ee
we thus obtain
\begin{cond} \label{ADcondition}
Affleck-Dine baryogenesis is thus only possible for 
\be
\mathbb{H}[Q,\Psi] \, < \, -1 \,\,.
\ee
\end{cond}

We note that the fact that conditions on soft masses can be expressed succinctly in terms of the holomorphic bisectional curvature was pointed out in \cite{Marsh:2011ud}, \cite{Dutta:2010sg}, where the formalism was applied to Affleck-Dine baryogenesis.

Taking the scale of inflation to be high, the conditions for inflation and baryogenesis are
\bea \label{ADandInf}
\mathbb{H}[\Psi]  & > & 0 \nonumber \\
\mathbb{H}[\Psi,Q] &\lesssim & -1 \,\,.
\eea

For situations where Eq.~ (\ref{constrelation}) holds, we have a clear contradiction and thus
\begin{nogo}
Affleck-Dine baryogenesis is intractable if the holomorphic sectional curvature at a given point of the Kahler manifold is constant for all choices of planes in the tangent space. In particular, this rules out Kahler manifolds with constant holomorphic sectional curvatures (in the above sense), and as a special case, maximally symmetric coset space.
\end{nogo}

There is no reason in principle for Affleck-Dine baryogenesis to be disallowed as long as Condition \ref{ADcondition} is met. However, for manifolds of the type mentioned above, the vacuum energy during A-D baryogenesis could not possibly also be driving slow-roll inflation. Also, as mentioned in the Introduction, these conclusions hold for $F-$term inflation; the case of $D-$term inflation is reserved for future work.


It is possible to reduce the condition for Affleck-Dine baryogenesis to a set of conditions on holomorphic sectional curvatures. Using Eq.~(\ref{ADandInf}) and Eq.~(\ref{associatedrelation}), it is clear that one requires
\begin{cond}
In terms of holomorphic sectional curvatures, the necessary condition for Affleck-Dine baryogenesis is
\be 
\sum_{a=1}^{4}\mathbb{H}[\lambda_a] - \mathbb{H}[Q]\, \lesssim \, - 4 \,\,,
\ee
\end{cond}
assuming that $\mathbb{H}[\Psi] \sim 0$. We will not apply this more general condition to any scenarios in this paper, leaving a more detailed exploration for a future publication.

\subsection{Examples}

As we have mentioned before, all spaces of constant holomorphic sectional curvature (in the sense that the holomorphic sectional curvature is isotropic in tangent space at all points in the manifold) are eliminated for viable Affleck-Dine baryogenesis, and in particular maximally coset manifolds are forbidden. This includes spaces with canonical Kahler potential
\be
K \, = \, \sum \Phi_{\ibar}\Phi_i 
\ee
and maximally symmetric coset spaces that appear in many string-inspired contexts
\be \label{Kahlercoset1}
K \, = \, -n \log (\Psi + \Psibar + Q \Qbar) \,\,\, \forall \, n \,\,.
\ee
Note that the constant holomorphic sectional curvature is given by $- 2/n$ and thus Affleck-Dine baryogenesis is by itself possible for $n \leq 2$, but that leaves inflation unviable. 

%
%

\section{Inflation and Supersymmetry Breaking} \label{infsusybreak}

In this Section, we discuss the case when the supersymmetry breaking during inflation persists as the supersymmetry breaking responsible for mass splitting in the visible sector. This scenario, shown in Eq.~(\ref{firstscenario}), is interesting for a variety of reasons mentioned in the Introduction.

Many things can be said about scenarios where the scales of inflation and supersymmetry breaking match, but perhaps the most far-reaching is that if both Nature prefers low-scale supersymmetry ($\sim \mathcal{O}(1)$) TeV, this necessitates models of low-scale inflation (for example \cite{German:2001tz}, \cite{Allahverdi:2009rm}). The main challenge then is to produce sufficient density perturbations, and this would typically require higher degrees of fine-tuning (in $\epsilon$) or extensions of the paradigm like the curvaton mechanism \cite{Lyth:2001nq}. It is interesting to ask if either of these options, that are typical only to low-scale inflation, can leave footprints on the mass splittings in the visible sector. For example, the extra fine-tuning of $\epsilon$ might introduce ingredients which affect the mediation mechanism of supersymmetry breaking. This was the avenue pursued in \cite{Allahverdi:2009rm}, where the Kahler modulus in a racetrack scenario in type IIB was taken as the inflaton, and the fine-tuning of $\epsilon$ affected the ratio of modulus to anomaly mediation.


One can go the other way, and ask what kinds of things the pattern of soft masses in the visible sector (apart from the overall scale) can tell us about the inflationary sector. Our focus in this work is on the geometry of the Kahler manifold in $D=4, N=1$ supergravity; one of the most important conditions on the geometry imposed by the soft masses is that of sequestering \cite{Randall:1998uk}.

The condition on the Kahler geometry for sequestering the soft masses amounts to
\be \label{sequesteringcond}
\mathbb{H}[\Psi,Q] \, = \, - \frac{1}{3}\,\,,
\ee
which sets the tree-level contribution to zero. This is evident from Eq.~\ref{firstscenario}. 

The condition for inflation is easily found by setting $\gamma = 1$ in Eq.~\ref{inflationfullcond}, corresponding to the fact that $H \sim m_{3/2}$. One finds $\mathbb{H}[\Psi] > - 1/3$, and thus from Eq.~(\ref{constrelation}), it is clear that 
\begin{cond}
If inflationary supersymmetry breaking and particle physics supersymmetry breaking are matched, sequestering is impossible on manifolds of constant holomorphic sectional curvature.
\end{cond}
In fact, for manifolds of constant holomorphic sectional curvature, the scalar masses satisfy a non-zero lower bound
\be \label{boundspaceform}
m^2_{i\overline{j}} \,\, \geq \,\, g_{i \overline{j}} \, m^2_{3/2} \, \frac{\gamma}{1+\gamma} \,\,. 
\ee

Of course, this does not exclude sequestering in such situations completely, but rather excludes only the most well-studied Kahler manifolds. For sequestering, Kahler potentials of the following form are studied
\bea
K  &=& -3 \log Y \nonumber \\
Y &=& Y(\Psi) \, + \, Y(Q) \,\,,
\eea
where $Y$ is separated between the supersymmetry breaking sector $\Psi$ and the visible sector field $Q$. Kahler potentials of this type always satisfy Eq.~(\ref{sequesteringcond}), as may be checked by direct computation. Only a subset of such manifolds are coset spaces, such as the Kahler potential in Eq.~(\ref{Kahlercoset1}), and these are excluded. 

Moreover, other simple forms of $Y$, such as 
\bea
Y \, &=& \, 1 - \sum \Phi_i \Phibar_{\ibar} \,\,\,\,\,\,\,\,\,\, {\rm or} \nonumber \\
Y \, &=& \, 1 -  2 \sum \Phi_i \Phibar_{\ibar} + \sum_{ij}(\Phi_i \Phibar_j)^2 
\eea
are excluded, since they also correspond to maximally symmetric coset spaces.

One could consider the next simplest forms of $Y$, which still satisfy the separability condition and hence the sequestering condition of Eq.~(\ref{sequesteringcond}), but do not correspond to maximally symmetric coset manifolds. For example, considering
\be
Y = (\Psi + \Psibar)^p - Q \Qbar
\ee
we obtain (as expected $\mathbb{H}[\Psi,Q] = -1/3$) and 
\bea
\mathbb{H}[\Psi]  =  \frac{2}{3} - \frac{1}{3p}(p-1)\{(p-2)(p-3)+ \nonumber \\
+(p-1)(p-2)^2 + p(p-1)^2 - 2p(p-1)(p-2)\} \,\,.
\eea
Clearly, even for $p=2$, we obtain $\mathbb{H}[\Psi] = -1/3$, which satisfies the necessary condition for inflation.

\section{Conclusion}

Supersymmetry, if it is a symmetry of Nature, is likely to have played a crucial role in the history of the very early universe. There are two features of supersymmetry that are especially important: the fact that the positive vacuum energy of the early universe breaks supersymmetry, and the fact that supersymmetric theories often contain flat directions. 

Both these features are relevant in Affleck-Dine baryogenesis. In this paper, we have probed the viability of Affleck-Dine baryogenesis from the point of view of the Kahler geometry of $D=4, N=1$ effective supergravity. In particular, we have looked at the initial condition problem of Affleck-Dine baryogenesis (the requirement that a flat direction acquire tachyonic soft mass due to inflationary supersymmetry breaking).

Moreover, and in a separate direction, it is interesting to pursue the idea that inflationary supersymmetry breaking and the breaking responsible for mass splittings in the visible sector are identified with each other. We have studied the question of sequestering in this context.

Our preferred tools for this study are the holomorphic sectional and bisectional curvatures of the Kahler manifold at a given point. The former is important because the requirement of small enough $\eta$ during inflation places a condition on the holomorphic sectional curvature of the supersymmetry breaking direction. The latter is important because the supersymmetry breaking-induced soft masses are most illuminatingly written in terms of the holomorphic bisectional curvature between the field breaking supersymmetry and the relevant visible sector field.

Relations between holomorphic sectional and bisectional curvatures can constrain the above physical scenarios. We have especially probed the simplest case when the holomorphic sectional curvature is isotropic in tangent space at a given point, resulting in a No-Go result for Affleck-Dine baryogenesis. The No-Go result extends to manifolds where such isotropy holds at all points, and in particular to maximally symmetric coset spaces.

It would be very interesting to probe the conditions on more general classes of manifolds, which we leave for future work. 

\section{Acknowledgements}

We would like to thank David Marsh and Claudio Scrucca for very helpful discussions, and comments on an earlier draft. This work is supported by DOE grant DE-FG02-95ER40917.

\appendix

\section{Sectional and Bisectional Curvatures}

The holomorphic bisectional curvature was first introduced by Goldberg and Kobayashi \cite{GoldbergKob}. In this Appendix, we give careful definitions of the various quantities used in the paper, and also derive the relations between sectional and bisectional curvatures that were used.

Let $\mathcal{M}$ be a Kahler manifold, and $R$ its Riemannian curvature tensor. At each point $x$ of $\mathcal{M}$, $R$ is a quadrilinear mapping 
\be
R : \,\, T_x(\mathcal{M}) \times T_x(\mathcal{M}) \times T_x(\mathcal{M}) \times T_x(\mathcal{M}) \longrightarrow \mathbb{R} \,\,.
\ee 

A plane $\sigma$ in $T_x(\mathcal{M})$ is said to be holomorphic if it is invariant by the almost complex structure $J$. Choosing an orthonormal basis $(X, JX)$ in $\sigma$, the holomorphic sectional curvature $B(\sigma)$ of $\sigma$ is defined as
\be
\mathbb{H}[\sigma] \, = \, R(X, JX, X, JX) \,\,.
\ee

Note that throughout the Appendix, we will use vectors in the uncomplexified basis, while in the text we used vectors in the complexified basis. 

For two $J-$invariant planes $\sigma$ and $\sigma^{\prime}$ in $T_x(\mathcal{M})$, and two unit vectors $X$ and $Y$ respectively on the planes, the holomorphic bisectional curvature is defined as
\be
\mathbb{H}[\sigma, \sigma^{\prime}] \, = \, R(X, JX, Y, JY) \,\,.
\ee

We now show the bound in Eq.~\ref{constrelation} for the case of manifolds of constant holomorphic sectional curvature, following an alternative route from Eq.~\ref{associatedrelation}.

For a Kahler metric $g$ of constant holomorphic sectional curvature $c$, the Riemann curvature is given by
\bea
&R&(X,Y,Z,W) \, = \, \frac{c}{4}[g(X,Z)g(Y,W) - g(X,W)g(Y,Z) + \nonumber \\
&+& g(X,JZ)g(Y,JW) - g(X,JW)g(Y,JZ) \nonumber \\
&+& 2g(X,JY)g(Z,JW) ] \,\,.
\eea
Then, considering two holomorphic planes $(X,JX)$ and $(Y,JY)$, using the Hermitian property of the metric, and the Bianchi identity
\be
R(X, JX, Y, JY) \, = \, R(X,Y,X,Y) + R(X,JY,X,JY) \,\,,
\ee
one obtains
\bea
R(X, JX, Y, JY) \, &=& \, \frac{c}{2}[g(X,X)g(Y,Y) + g(X,Y)^2 \nonumber \\
&+& g(X,JY)^2 ] \,\,.
\eea

Clearly,
\be
\frac{c}{2} \, \leq \, \mathbb{H}[\sigma, \sigma^{\prime}] \, \leq \, c \,\,.
\ee
The lower limit is obtained when $(X,JX)$ and $(Y, JY)$ are orthogonal. This gives the required bound, in a manner alternative from Eq.~\ref{associatedrelation}.


\subsection{General Relations}

On general manifolds, the holomorphic bisectional curvature can be expressed in terms of the holomorphic sectional curvatures of certain holomorphic and anti-holomorphic sections associated with the section spanned by the pair $(X, Y)$ at any given point $x \in \mathcal{M}$.
\bea \label{genappendix}
R(X,JX,Y,JY) = \frac{1}{4} (\mathbb{H}[X+Y] + \mathbb{H}[X-Y] \nonumber \\
+ \mathbb{H}[X+JY] + \mathbb{H}[X-JY] -\mathbb{H}[X] -\mathbb{H}[Y] ) \,\,.
\eea

The proof follows from the relation
\bea
\mathbb{H}[X+Y] + \mathbb{H}[X-Y] = \frac{1}{2}(\mathbb{H}[X] \nonumber \\
+ \mathbb{H}[Y] +6\mathbb{H}[X,Y] - 4K(X,Y))
\eea
and the corresponding relation with $Y$ replaced by $JY$, and then the use of the first Bianchi identity. 

The relation Eq.~\ref{genappendix} holds for $X$ and $JY$ orthonormal. In more general cases, $<X, JY> = \rm{cos} \theta$ appears in the expression. The equation is particularly useful if the holomorphic sectional curvatures of a manifold $\mathcal{M}$ can be bounded, as happens in holomorphically pinched manifolds where $\lambda \leq \mathbb{H}[X] \leq 1$. On such manifolds, the bisectional curvature becomes bounded as well.

%
%

\end{document}